\documentclass[twocolumn,prl,aps,showpacs,superscriptaddress]{revtex4}
\usepackage{amsmath}
\usepackage[psamsfonts]{amssymb}
\usepackage{bm}
\usepackage{dcolumn}
\usepackage{color}

\usepackage{units}

\usepackage{ifpdf}
\ifpdf
\usepackage[pdftex, final]{graphicx}
\else
\usepackage[dvips, final]{graphicx}
\fi                                     

\usepackage{setspace}
\newcounter{NBc}

\begin{document}

\title{Heavy Ion-Beam Driven Isentropic Compression
Experiments}

\author{A.~Grinenko}
\author{D.O.~Gericke}
\affiliation{Centre for Fusion, Space and Astrophysics,
             Department of Physics, University of Warwick,
             Coventry CV4 7AL, United Kingdom}

\author{D.~Varentsov}
\affiliation{GSI Helmholtzzentrum f\"ur Schwerionenforschung GmbH,
	     Planckstr. 1, 
	     64291 Darmstadt, Germany}

\begin{abstract}
A new design for heavy-ion beam driven isentropic compression experiments is
suggested and analysed. The proposed setup utilises the long stopping ranges and the
variable focal spot geometry of the high-energy uranium beams delivered at the
GSI Helmholtzzentrum f\"ur Schwerionenforschung and Facility for Antiproton and
Ion Research accelerator centers in Darmstadt, Germany, to produce a planar ramp
loading of various samples. In such experiments, the predicted high pressure
amplitudes (up to~\unit[10]{Mbar}) and short timescales of compression
(below \unit[10]{ns}) will allow to test the time dependent material deformation
phenomena at unprecedented extreme conditions.
\end{abstract}


\maketitle

Research carried out in astrophysics, planetary and material sciences seek a thorough understanding of the behavior of matter at high pressures. For example, the equation of state (EOS) of iron under pressures of~\unit[1--4]{Mbar} is crucial in order to determine the state of the Earth’s core~\cite{laio00, belonoshko03, boehler00}. EOS data around~\unit[0.1]{Mbar} is required to establish the state and the composition in Earth’s lower mantle~\cite{boehler00, zhang99}, while the dynamics of the processes in the mantle is considered to be dependent on the structural phase transformation kinetics~\cite{solomatov94}. The pursuit after materials for technological applications~\cite{mcmilan02} also entails a detailed understanding of the kinetics of the high-pressure phase transitions. Modelling the physical processes during the projectile impact relevant to meteoroid protection and crater formation~\cite{okeefe99} requires the dynamical response of solids at ultrahigh strain rates. Hydrogen EOS at high pressures are especially important for understanding the structure and evolution of the hydrogen-bearing astrophysical objects such as the giant planets like Saturn and Jupiter~\cite{guillot99,sauman00} as well as for the inertial fusion energy research~\cite{azteni04,lindl04}.

\begin{figure}[b]
\includegraphics[width = 6 cm,clip=true]{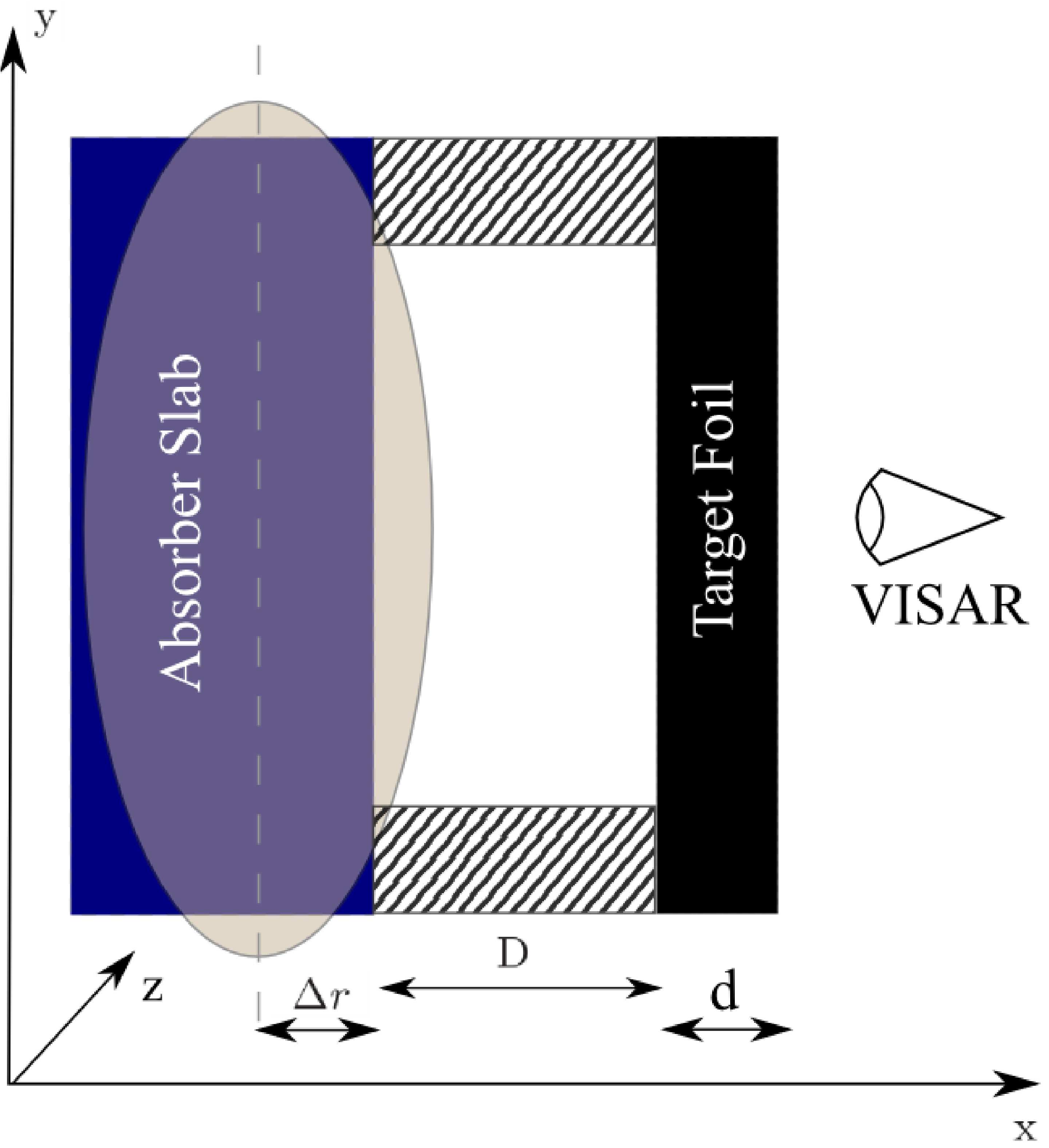}
\caption{Schematic of the experimental design. The ion beam with an elliptic focal
spot propagates in $z$ direction. It heats the absorber slab along its surface.
The depth of the beam inside the absorber, i.e. the distance between the
beam's axis and the inner surface of the slab is $\Delta r.$ The target foil
of thickness $d$ is placed parallel to the absorber at a distance $D$.
Hatched areas represent the support washer used to and to coax the expanding
absorber material. The velocity of the rear target surface is recorded using a
line-imaging VISAR.\label{Fig:sketch}}
\end{figure}

The technique of isentropic compression using ramp wave loading (RWL)~\cite{reisman00, hayes01, cauble02, hayes04, rothman05, davis06, chhabildas86, barnes74, edwards04, lorentz05, swift05, smith07} allows one to sample  EOS data along an isentrope up to several Mbar pressures, located on the phase diagram between the parameter regions accessible in diamond anvil cell~\cite{weir59,jarayama86} and shock wave~\cite{mcqueen84,neal76} experiments. Unlike a shock wave experiment where a single point on a shock adiabat is obtained, in isentropic compression experiments a continuous set of data points is recorded and the solid state of a sample is ensured up to high pressures. RWL technique was also shown to be a more sensitive tool for studying the dynamics of ultrafast structural phase transformations than the shock-wave based~\cite{smith08, bastea05, dolan07} methods.

\begin{table}[t]
 \caption{Parameters of the experiments with Al [cases (a)-(d)] and Fe
	[cases (e)-(h)] targets (see Fig.~\ref{Fig:sketch}). $E_0$ is the
	ion energy, $N_0$ is the number of ions per pulse, $\tau$
	is the pulse duration, and $\Delta x \times \Delta y$ are the FWHM
	dimensions of the beam focal spot. The depth of the beam in 	the
	absorber is $\Delta r = 0.2\,\rm mm$ in all the cases. All spatial
	dimensions listed are in millimeters.
\label{tab:parameters_Al}}
 \begin{ruledtabular}
 \begin{tabular}{ccccccc}

  & $E_0$ [GeV/u] & $N_0$& $\tau$ [ns]& $\Delta x \times \Delta y $ &
     D & d\\ 
  \hline
  (a)& 2.7 & $10^{12}$ & 50 & $0.3\times0.5$ & 0.2 & 0.1\\
  (b)& 2.7 & $10^{12}$ & 100 & $0.3\times1.0$ & 0.5 & 0.1\\
  (c)& 0.2 & $10^{11}$ & 100 & $0.3\times1.0$  & 0.5 & 0.2\\
  (d)& 0.35 & $10^{10}$ & 100 & $0.3\times0.5$  & 0.2 & 0.2\\
  (e)& 2.7 & $10^{12}$ & 50 & $0.3\times0.5$ & 0.2 & 0.1\\
  (f)& 2.7 & $10^{12}$ & 50 & $0.3\times0.5$ & 0.2 & 0.05\\
  (g)& 2.7 & $10^{11}$ & 100 & $0.3\times1.0$  & 0.5 & 0.1\\
  (h)& 0.2 & $10^{10}$ & 50 & $0.3\times1.0$  & 0.5 & 0.075\\
 \end{tabular}
 \end{ruledtabular}
\end{table}

RWL has been demonstrated with different drivers, such as magnetic pulse loading using high-current pulsed power generators~\cite{reisman00, hayes01, cauble02, hayes04, rothman05, davis06}, high-power lasers~\cite{edwards04, lorentz05, swift05, smith07}, and gas guns~\cite{chhabildas86} or high explosives~\cite{barnes74} using graded density impactors. The typical loading times with these drivers are \unit[10]{ns}, \unit[100]{ns} and \unit[1]{$\mu$s}, respectively. In this Letter, a new scheme for planar isentropic compression experiments using an intense heavy ion beam as a driver is proposed and analysed. The long absorption range and the variable focal spot size and shape of energetic ion beams allows one to design isentropic compression experiments with fairly planar geometry. The beam parameters needed to generate pressures of up to~\unit[4]{Mbar} in aluminium and~\unit[7]{Mbar} in iron are well within the reach of the uranium beams that will be delivered at the new Facility for Antiproton and Ion Research (FAIR) which is being built in Darmstadt. The beam intensities needed to generate pressures approaching~\unit[1]{Mbar} shall become available with the completion of the high-current upgrade of the SIS-18 heavy ion synchrotron of the GSI Helmholtzzentrum f\"ur Schwerionenforschung in Darmstadt~\cite{spiller06} in the year 2009. In RWL experiments with intense ion beams, the typical compression times from zero to the maximum pressure are about~\unit[20]{ns} for the aluminium and about~\unit[10]{ns} for the iron samples considered here, which is comparable with the dissipative relaxation times of aluminium and iron~\cite{swegle85, boettger97}. Resent laser-driven experiments~\cite{smith07} have demonstrated that at such short time scales there exists a stiffer response than had been expected from previous slower ramp compression experiments and from models based on either static or shock-wave experiments.

The layout of the experiment is shown in Fig.~\ref{Fig:sketch}. The ion beam with an elliptical focal spot and  Gaussian transverse intensity distribution heats an absorber slab along its surface. Lead is chosen as the absorber material. Since the stopping ranges of energetic $^{238}$U ions in lead absorber are much larger than the length of the absorber slab, the latter is heated uniformly along the $z$ axis. The focal spot size and its aspect ratio can be varied, what allows for generation of a quasi-planar ramp wave. In the similar manner as in the laser-driven ramp compression or shock wave experiments~\cite{smith07}, a stepped target with rather wide steps along $z$ direction can be used. 

\begin{figure}[b]
\includegraphics[width = \columnwidth,clip=true]{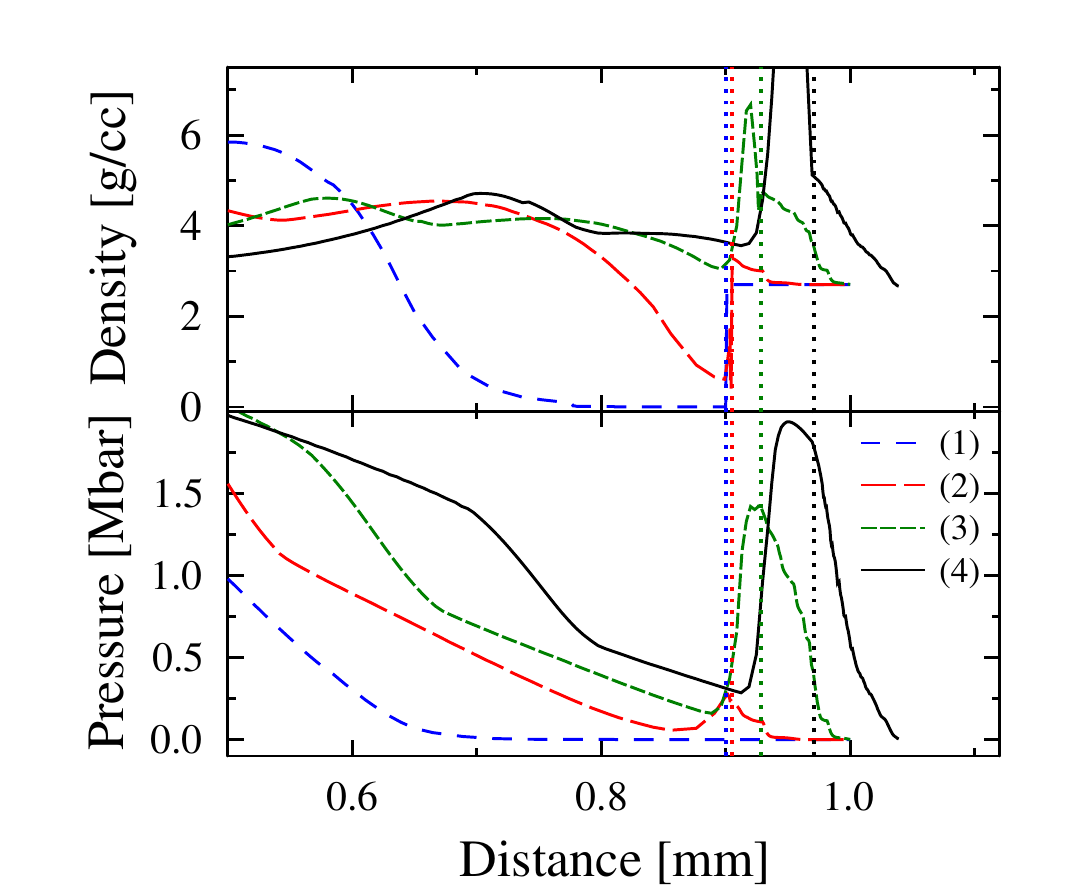}
\caption{ Evolution of pressure and density for the case (b) of
Table~\ref{tab:parameters_Al}. The dotted vertical lines indicate the
location of the target front surface at different times. The phases of
the process are: (1) $t = 59\,\rm ns$~--- the evaporated absorber material~---
lead is accumulating at the front surface of the aluminium sample; (2) $t =
78\,\rm ns$ --- the isentropic compression has started; (3) $t = 86\,\rm ns$ ---
the accumulation continues; the compression wave reaches the rear surface of the
sample before the shock wave has formed; (4) $t = 94\,\rm ns$ --- the motion of
the rear surface of the target foil can be detected. 
\label{Fig:Al_p_t}}
\end{figure}

Similarly to laser-driven RWL experiments~\cite{edwards04}, the process of
the ion beam driven RWL consists of three distinct phases. First, the ion beam
evaporates the absorber material off its surface or unloads the material due to
a shock wave generated inside the absorber, depending on the beam intensity,
focal spot size and the distance $\Delta r$ between the beam's axis and the
absorber surface. At the next stage, the absorber's material expands into the
vacuum gap. The adjustable displacement of the beam axis relative to the
absorber surface allows for some degree of control over the plasma expansion regime.
In the last stage, the absorber's plasma piles up against the target foil thus
producing a smoothly increasing pressure load in the sample. The evolution of
the compression wave launched into the sample is shown in Fig.~\ref{Fig:Al_p_t}.
One can see that the compression wave gradually steepens with the distance
traveled before it breaks into a shock wave. This distance determines the
maximal allowed thickness of the target foil.

Provided the compression wave launched into the sample does not break into a
shock wave, the pressure and density history at the front surface of the sample
can be reconstructed using the bootstrap back integration method~\cite{hayes01}.
The method uses the rear surface velocity history as an initial condition for a
Riemann solver.

The major benefit of the proposed experimental design as compared to other approaches for EOS measurement with heavy ion beams~\cite{arnold82, hoffmann02} is twofold. Firstly, the EOS along a compression adiabat can be determined by measuring only one parameter~--- the rear surface velocity employing a line-imaging VISAR as the principal diagnostics. Secondly, this design does not rely on detailed knowledge of the beam-matter interaction processes like the stopping power, since the beam energy is not deposited into the sample directly \cite{grinenko08:PRL} but is converted to the kinetic energy of the absorber material. In order to interpret and use the results of the experiment one therefore does not have to precisely measure the transverse distribution of the beam intensity at the focal plane, which can be problematic for intense focused heavy ion beams~\cite{varent08}. Moreover, the presented design has also an important advantage that the target is not being preheated by energetic secondary particles and projectile fragments produced during the interaction of the beam with the absorber material. Furthermore, this approach requires neither high accuracy of the beam-target alignment, nor good short-to-shot reproducibility of the beam parameters.

\begin{figure}[t]
\includegraphics[width = \columnwidth]{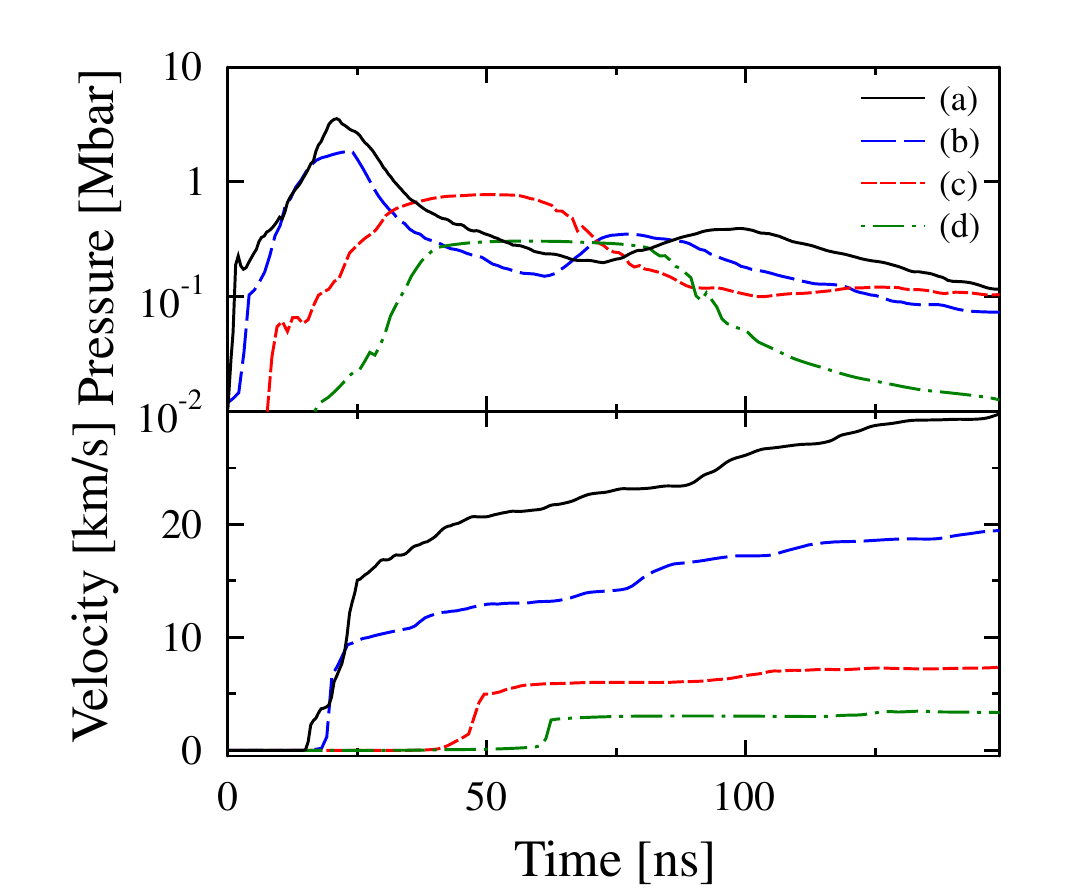}
\caption{Rear surface velocity and front surface pressure histories for
different cases described in Table~\ref{tab:parameters_Al}.}
\label{Fig:Al_v_p}
\end{figure}

Parameters of the beam and the target geometries considered in this work are summarised in Table~\ref{tab:parameters_Al}. The cases (a),(e) and (b),(f) in the table correspond to the beam parameters of the SIS-100 synchrotron to be available at FAIR~\cite{FBTR}, the cases (c),(g) correspond to the beams that will be available after completion of the SIS-18 upgrade~\cite{spiller06}, and the cases (d),(h) correspond to the present SIS-18 beam available for high energy density physics experiments~\cite{varent08,varent06}. A lead absorber slab of~\unit[400]{$\mu$m} thickness was considered for all the analysed cases. The geometric parameters of the target have been adjusted to guarantee the planarity of the ramp compression wave propagating in the $x$ direction for at least~\unit[$y \leq \pm 50$]{$\mu$m}, and to ensure a shockless compression of the target. The planarity of the designed experiment justifies a one dimensional analysis using the 1D~hydrodynamic simulation code~\cite{grinenko09}.

The calculated histories of the pressure at the front and the velocity at the rear surfaces of the aluminium samples are shown in Fig.~\ref{Fig:Al_v_p}. The planarity of the compression wave can be adjusted by varying the ion beam extension in the $y$ direction. However, this also affects the level of specific energy density deposited by the beam in the absorber. Two possible beam cross sections are compared: in the case (a) the FWHM height of the beam spot is \unit[0.5]{mm}, whereas in the case (b)~--- \unit[1]{mm}. Increasing the height of the beam spot allows one to increase the vacuum gap which enhances the smoothness of the loading without loosing in the planarity. The curves corresponding to these two cases indicate that the enhancement of the smoothness of the loading compromises the amplitude of the compression: the maximum pressure corresponding to case (a) is about \unit[3.5]{Mbar}, whereas in case (b) it is only about \unit[2]{Mbar}. The maximum pressure that one can obtain using upgraded SIS-18 beam (c) is about \unit[1]{Mbar} and the currently available SIS-18 beam (d) can isentropically load aluminium up to \unit[500]{kbar}.

The strain rate is approximately $8\times 10^7$ $\rm s^{-1}$ at the peak pressure of about \unit[2]{Mbar} in the examined case (b) (see Table~\ref{tab:parameters_Al}). The corresponding time to the maximum compression is approximately \unit[20]{ns}, which is comparable to the rise time of low-stress steady shock~\cite{swegle85}. The laser RWL experiments by Smith {\em et al.}~\cite{smith07} at similar compression times and strain rates have demonstrated a stiffening of the stress-strain response. Using ion beams, the adjustable ion pulse duration, focal spot size and the depth of the ion beam in the absorber allow a degree of control over the compression time, amplitude and strain rate. This provides a unique tool to carry out parametric studies of the stress-strain response.

\begin{figure}[t]
\includegraphics[width = 8.0cm,clip=true]{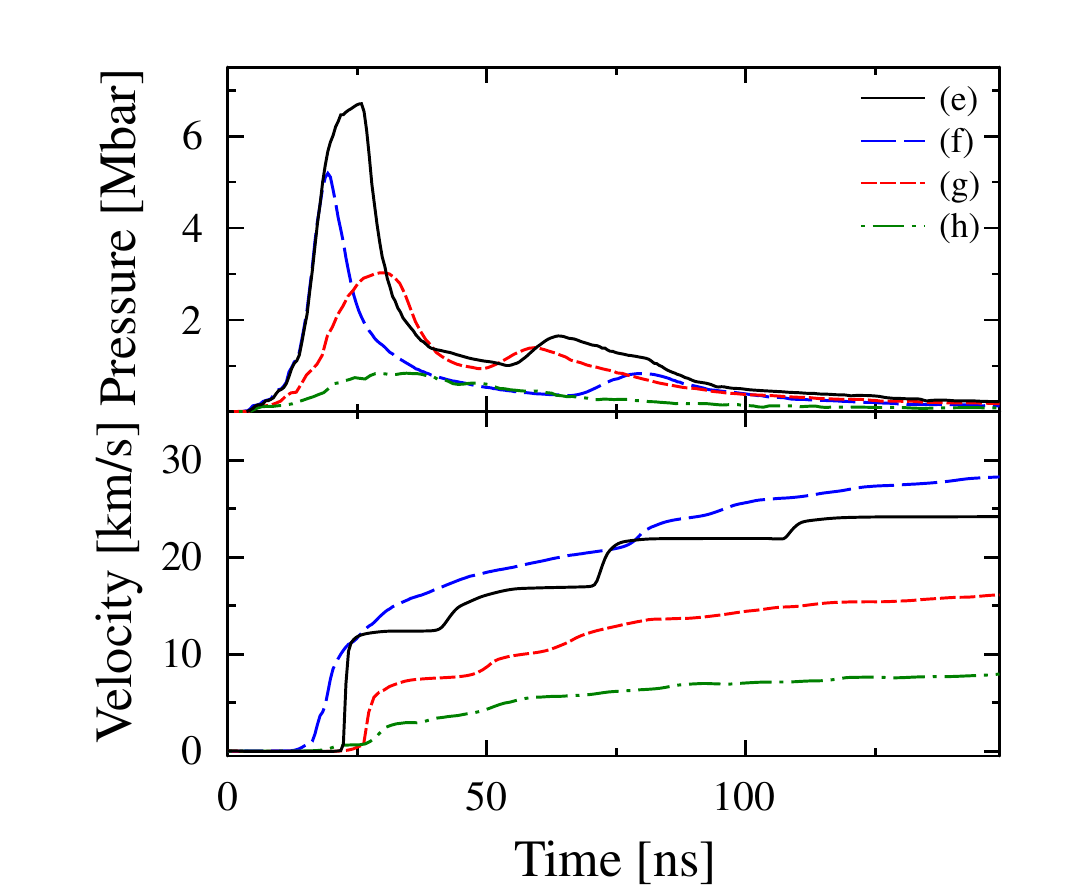}
\caption{Rear surface velocity and front surface pressure histories for
	different cases of Fe samples described in 
	Table~\ref{tab:parameters_Al}.}
\label{Fig:Fe_v_p}
\end{figure}

A similar analysis has been carried out for iron samples (Table~\ref{tab:parameters_Al}). The corresponding histories of the front face pressure and rear face velocity are shown in Fig.~\ref{Fig:Fe_v_p}. One can see that a {Mbar} pressure loading is within the rich of the existing SIS-18 beam, whereas the upgrade of the machine can provide the compression data of up to about~\unit[3]{Mbar}  in the nearest future. Up to~\unit[6]{Mbar} pressures will be generated in the corresponding experiments at FAIR with extremely high strain rates of the order of $10^8\,\rm s^{-1}.$

The curves (e) and (f) in Fig.~\ref{Fig:Fe_v_p} demonstrate the differences between the velocity traces in case of shock and adiabatic compression, respectively. In order to ensure shockless loading, the sample's thickness had to be reduced from~\unit[100]{$\mu$m} as in the case (e) to~\unit[50]{$\mu$m} as in the case (f). Obviously, decreasing the sample's thickness causes the motion of the sample's rear surface to start earlier. This motion acts to release the pressure of the absorber's material piling up at the front surface of the sample and to reduce the pressure amplitude in the target. Therefore, a compromise should be devised to ensure the adiabatic character of the compression on one side, and to obtain the highest possible pressure on the other.

The capabilities of the high-energy heavy ion beam accelerators available at GSI and later at FAIR to drive planar isentropic compression of solids were analysed. It was shown that the time scales of the loading are comparable with the fastest laser drives~\cite{edwards04,smith07}, and the amplitudes of pressure surpass those obtained using the high-power magnetic drives~\cite{davis06}. These features allow to investigate the dynamical response of solids in the regime of previously unattainable parameters and possibly the dynamics of the pressure-induced structural phase transformations~\cite{smith08,bastea05,dolan07}. Moreover, ramp loading with pressures higher than \unit[0.1]Mbar can be obtained in multy-layer targets~\cite{dolan07} using the currently existing SIS-18 ring for the purpose of studying the pressure induced phase transformations in water. 

Support from EPSRC and INTAS is gratefully acknowledged.

\bibliographystyle{myapsrev}
\bibliography{ice}

\end{document}